\documentclass[aps,prb,a4paper,twocolumn,floatfix,showpacs,preprintnumbers]{revtex4-1}
\usepackage{amsmath}
\usepackage{amssymb}
\usepackage{graphicx}
\usepackage{color}

\begin{document}

\title{Conductivity and scattering in graphene bilayers: numerically exact
results vs. Boltzmann approach }
\author{Hengyi Xu}\affiliation{Condensed Matter Physics Laboratory, Heinrich-Heine-Universit\"at,
Universit\"atsstr.1, 40225 D\"usseldorf, Germany}

\author{T. Heinzel}\affiliation{Condensed Matter Physics Laboratory, Heinrich-Heine-Universit\"at,
Universit\"atsstr.1, 40225 D\"usseldorf, Germany}
\author{I. V. Zozoulenko}
\email{igor.zozoulenko@itn.liu.se}\affiliation{Solid State Electronics, Department of Science and Technology (ITN), Link\"{o}ping
University, 60174 Norrk\"{o}ping, Sweden}

\date{\today }

\begin{abstract}
We derive analytical expressions for the conductivity of bilayer graphene
(BLG) using the Boltzmann approach within the the Born approximation for a
model of Gaussian disorders describing both short- and long-range impurity
scattering. The range of validity of the Born approximation is established
by comparing the analytical results to exact tight-binding numerical
calculations. A comparison of the obtained density dependencies of the
conductivity with experimental data shows that the BLG samples investigated
experimentally so far are in the quantum scattering regime where the Fermi
wavelength exceeds the effective impurity range. In this regime both short-
and long-range scattering lead to the same linear density dependence of the
conductivity. Our calculations imply that bilayer and single layer graphene
have the same scattering mechanisms. We also provide an upper limit for the
effective, density dependent spatial extension of the scatterers present in
the experiments.
\end{abstract}

\maketitle

\section{Introduction.}

Single- and bilayer graphene have attracted tremendous attention since their
discovery \cite{Novoselov2004} due to their fascinating and in many respects
unique properties (for a review see, e.g., Refs. %
\onlinecite{CastroNetoreview,DasSarmareview,Peresreview}). Presently, one of
the central issues in graphene research is to identify the scatterers that
dominate the conductivity $\sigma $. This is not only a fundamental
question, but also the prerequisite for progress regarding the quality of
the electronic properties of graphene devices. Since the character of the
scatterers manifests itself directly in the electron density ($n$)
dependence of $\sigma $, this function plays a key role in the corresponding
experimental and theoretical investigations. \cite%
{DasSarmareview,Peresreview} Most experiments in single-layer graphene (SLG)
show a dependence close to $\sigma \propto n$. \cite%
{Novoselov2005,Tan2007,Morozov2008,Peresreview}, while some sub-micrometer
extra-clean suspended samples show a square-root dependence $\sigma \propto
n^{1/2}$ consistent with the ballistic transport regime\cite%
{Du,BolotinPRL,Klos2009}. Within the standard Boltzmann transport theory in
the Born approximation, \cite{Ando2006,Nomura2006,Hwang2007} the linear
density dependence emerges from long-range Coulomb scattering which implies
that charged impurities in the substrate are the dominant scatterers.
Deviations from a linear dependence were attributed to effects of short
range impurities\cite{Novoselov2005,Tan2007,Morozov2008,Jang2008,Zhu2009},
for which the Boltzmann theory predicts $\sigma =$ const \cite%
{Ando2006,Nomura2006,Hwang2007}. This point of view has been challenged by
theories going beyond the first Born approximation\cite%
{Stauber2007,KatsnelsonNovoselov,Ostrovsky} as well as numerical simulations%
\cite{Klos2010,Wehling2010} which show that short-range disorder leads to a
linear density dependency similar to the one for the long-range potential.
At the same time, a growing evidence from recent experiments\cite%
{Ponomarenko2009,Monteverde2010,Ni2010} points to short-range impurities as
the main scattering mechanism in SLG.

The situation is much less clear in bilayer graphene (BLG). In most studies
the BLG conductivity exhibits a linear density dependence very similar to
that one of SLG. \cite{Morozov2008, Monteverde2010,Xiao2010} A superlinear
dependence was reported by Zhu \textit{et al.}\cite{Zhu2009}, whereas a
sublinear behavior has been observed in sub-micrometer BLG samples\cite%
{Feldman2009} which may be an indication quasi-ballistic transport. In the
density regime where these experiments have been carried out, the energy
dispersion of BLG is parabolic. For the case of short-range disorder, both
the standard Boltzmann approach\cite{Adam2008,DasSarma2010}, as well as
theories not relying on the Born approximation\cite{Katsnelson2007,Ferreira}
predict $\sigma \propto n$. Recent numerical modeling of the BLG
conductivity based on the Kubo approach shows that sufficiently far away
from the neutrality point, a linear or sublinear dependence of $\sigma $ on $%
n$ is found, depending on the strength and type of of the disorder.\cite%
{Katsnelson2010,Trushin} As far as the long-range disorder is concerned, the
Boltzmann model in the Born approximation predicts $\sigma \propto n^{2}$
for unscreened, $\sigma \propto n$ for overscreened and $\sigma \propto
n^{\gamma }$ with $1<\gamma <2$ for screened Coulomb disorder \cite%
{DasSarma2010}. These predictions are qualitatively different from the
corresponding predictions for SLG.\cite%
{Ando2006,Nomura2006,Hwang2007,Peresreview} Therefore, in order to explain
the observed density dependence (which is very similar for SLG and BLG), Das
Sarma \textit{et al.}\cite{DasSarma2010} concluded that scattering
mechanisms are fundamentally different for SGL and BLG, being dominated by
the Coulomb impurities for the former and short-range impurities for the
later. This is a rather surprising conclusion because scattering mechanisms
in both systems are expected to be the same as both SLG and BLG are produced
by the same manufacturing technique with the same substrate used in the
measurements.

Since conclusions regarding the nature of the scattering mechanism in BLG
are often based on the predictions of the standard Boltzmann approach, it is
of utmost importance to establish its range of validity and to study how
well it models the exact conductivity. In the present study, we utilize the
well established model of Gaussian disorder, \cite%
{Bardarson,Lewenkopf,Adam2009,Klos2009,Klos2010,Peresreview} where the
effective screening length $\xi $ can be chosen to describe both short- and
long-range scattering, and derive analytical expressions for the
conductivity of the BLG within the standard Boltzmann approach in the Born
approximation. We then perform exact numerical tight-binding (TB)
Landauer-type calculations and compare them with the analytical results. We
demonstrate that for the density regime corresponding to the parabolic
dispersion the exact TB calculations are consistent with the Boltzmann
predictions. We show that in the experimentally relevant regime the latter
predicts a linear density dependence of the conductivity for both short- and
long range scattering\textbf{.} This density dependence is in agreement with
most experiments reported to date. Our calculations imply that the same
scattering mechanisms dominate in BLG and SLG. We also provide an estimate
for the range of effective screening lengths for the scatterers present in
the experiments.

\section{Description of the model.}

We calculate $\sigma (n)$ of BLG using the standard $p$-orbital
nearest-neighbor tight-binding Hamiltonian

\begin{eqnarray}
H &=&\sum_{\ell ,\langle i,j\rangle }(V_{i}a_{\ell ,i}^{+}a_{\ell
,i}+V_{j}b_{\ell ,j}^{+}b_{\ell ,j})  \label{H} \\
&&-t\sum_{\ell ,\langle i\rangle }(a_{\ell ,i}^{+}b_{\ell ,i+\Delta
}+h.c.)-t_{\perp }\sum_{i}(a_{1,i}^{+}b_{2,i}+h.c.),  \notag
\end{eqnarray}%
with $V_{i}$ being the external potential at site $i,$ $a_{\ell ,i}^{+}$ ($%
b_{\ell ,j}^{+}$) being the creation operator at sublattice $A$ ($B$) and
site $i(j)$ in layer $\ell =1,2$, and the coupling integrals $t=3.16$ $%
\mathrm{eV,}$ $t_{\perp }=0.39$ $\mathrm{eV.}$ \cite%
{CastroNetoreview,DasSarmareview,Peresreview} In the second term the index $%
i+\Delta $ corresponds to the nearest neighbors to the site $i.$ In the
parabolic dispersion regime we follow Yuan \textit{et al.}\cite%
{Katsnelson2010} and set $t_{\perp }=0.5t$. This effectively extends the
parabolic band to higher densities. Without this scaling and for the
impurity concentrations used in our numerical calculations, the parabolic
band would otherwise not be accessible since it is essentially governed by
the impurity band. Scattering centers are modeled by the Gaussian potential,%
\cite{Bardarson,Lewenkopf,Adam2009,Klos2009,Klos2010,Peresreview}%
\begin{equation}
V(\mathbf{r}_{i})=\sum_{i^{\prime }=1}^{N_{imp}}U_{i^{\prime }}\exp \left( -%
\frac{|\mathbf{r}{_{i}}-\mathbf{r}_{i^{\prime }}|^{2}}{2\xi ^{2}}\right) .
\label{Gaussian}
\end{equation}%
We refer to the case $\xi =a$ (appropriate for the absorbed neutral
impurities) as short-range scattering, and to the case $\xi \gg a$
(appropriate for the remote charged impurities) as long-range scattering,
with $a$ being the C-C distance. The potential heights are uniformly
distributed in the range $U_{i}\in \lbrack -\delta ,\delta ]$ where $\delta $
denotes the maximum potential height. The correlator of this potential has
also a Gaussian shape, \cite{Bardarson,Lewenkopf,Adam2009}
\begin{equation}
C(r)\equiv \sum_{i}V(\mathbf{r}_{i})V(\mathbf{r}_{i}-\mathbf{r})=\frac{%
K(\hbar v_{F})^{2}}{2\pi \xi ^{2}}\exp \left( -\frac{r^{2}}{2\xi ^{2}}%
\right) ,  \label{correlator}
\end{equation}%
where the dimensionless impurity strength is described by the parameter $%
K\approx 40.5n_{imp}(\delta /t)^{2}(\xi /\sqrt{3}a)^{4}$ with $n_{imp}$
being the relative concentration of impurities.

For the exact numerical calculations, we consider a rectangular BLG stripe
of length $L$ and width $W$ exposed to the impurity potential according to
Eq. (\ref{Gaussian}) and attach it to semi-infinite leads. The conductivity
of the stripe $\sigma (n)=\frac{L}{W}G(n)$ as a function of the electron
density $n$ is obtained from the transmission coefficient $T$ which is
related to the conductance $G$ via the Landauer formula $G=\frac{2e^{2}}{h}%
T. $ $T$ and $n$ are computed with the aid of the recursive Green's function
technique \cite{Xu2008,Xu2009}. Because of computational limitations we
study stripes with $L/W>1$. The obtained results are insensitive to $L/W$ as
long as $L/W>1$.

\section{The Boltzmann approach}

Within the Boltzmann approach, the conductivity reads\cite{Ashcroft}

\begin{equation}
\sigma ={\frac{1}{2}}e^{2}\tau D(E_{F})\left( v_{F}^{BG}\right) ^{2},
\label{sigma}
\end{equation}%
where $D(E_{F})$ is the density of states (DOS) at the Fermi energy, $%
v_{F}^{BG}$ is the Fermi velocity of the bilayer graphene, and $\tau $ is
the scattering time. An analytical expression for the Boltzmann conductivity
for the bilayer graphene can be obtained for the limiting cases of the
parabolic and linear dispersions, $\hbar v_{F}|k|\ll \frac{t_{\perp }}{2}$
and $\hbar v_{F}|k|\gg \frac{t_{\perp }}{2}$ respectively (or, equivalently,
$n\ll n_{0}$ and $n\gg n${}$_{0},$ where $n_{0}=\frac{1}{\pi (3a)^{2}}\left(
\frac{t_{\perp }}{t}\right) ^{2}$ is the critical density separating the
parabolic and linear bands, and $v_{F}=\frac{3at}{2\hbar }$ is the Fermi
velocity of single-layer graphene).

In this section we present analytical expressions for the conductivity $%
\sigma $ for the bilayer graphene. The details of the derivation are given
in the Appendix. For the parabolic dispersion ($n\ll n_{0}$) the
conductivity reads,
\begin{eqnarray}
\sigma &=&\frac{e^{2}\hbar ^{3}}{m^{\ast 2}}\frac{n}{K(\hbar v_{F})^{2}}%
\frac{e^{z}}{\left( \frac{1}{z}-1\right) I_{1}(z)+\left( \frac{1}{z}%
+1\right) I_{2}(z)}  \label{sigma_par} \\
&\propto &\left\{
\begin{array}{c}
n,\;z\ll 1\text{ (quantum scattering);} \\
n^{5/2},\;z\gg 1\text{ (classical scattering).}%
\end{array}%
\right.  \notag
\end{eqnarray}%
where $I_{\nu }$ is the modified Bessel function, and $z=\pi n\xi
^{2}=\left( \frac{2\pi \xi }{\lambda }\right) ^{2}$ with $\lambda $ being
the Fermi wavelength. It follows from the definition of $z$ that the
condition $z\ll 1$ corresponds to the case of quantum scattering when the
Fermi wavelength is larger than the effective width of the potential
barrier, $\lambda \gg \xi $, while the opposite condition $z\gg 1$
corresponds to the case of classical scattering, $\lambda \ll \xi $.

For the linear dispersion ($n\gg n_{0})$ the conductivity reads,
\begin{eqnarray}
\sigma &=&\frac{16e^{2}}{h}\frac{ze^{z}}{K\left( I_{1}(z)+I_{2}(z)\right) }
\label{sigma_lin} \\
&\propto &\left\{
\begin{array}{c}
\text{const},\;z\ll 1\text{ (quantum scattering);} \\
n^{3/2},\;z\gg 1\text{ (classical scattering).}%
\end{array}%
\right.  \notag
\end{eqnarray}

\begin{table}[t]
\begin{tabular}{|l|l|l|}
\hline
& parabolic band & linear band \\
& ($n\ll n_{0}$) & ($n \gg n_{0}$) \\ \hline
\textrm{quantum scattering} & Regime I: & Regime II: \\
$(z\ll 1)$ & $\sigma \propto n$ & $\sigma $ = const \\ \hline
\textrm{classical scattering} & Regime III: & Regime IV: \\
$(z\gg 1)$ & $\sigma \propto n^{5/2}$ & $\sigma \propto n^{3/2}$ \\ \hline
\end{tabular}%
\caption{Four different regimes for the density dependence for the
conductivity of BLG as predicted by the Boltzmann theory, Eqns. (\protect\ref%
{sigma_par}),(\protect\ref{sigma_lin}). It is noteworthy that the exponents
for the linear band of BLG (regimes II and IV) are the same as those for SLG%
\protect\cite{Adam2009,Klos2010}. }
\end{table}
Hence, it follows from Eqs. (\ref{sigma_par}) and (\ref{sigma_lin}) that the
Boltzmann approach for the Gaussian potential predicts four different
regimes where the density dependence of the conductivity $\sigma =\sigma (n)$
is qualitatively different for parabolic and linear bands ($n\ll n_{0}$
respectively $n\gg n${}$_{0}$) and for quantum and classical scattering ( $%
z\ll 1$ respectively $z\gg 1$). The corresponding asymptotes for the density
dependence for these four regimes are summarized in Table I.

\section{Exact results and discussion}

\begin{figure}[tbp]
\begin{centering}
\includegraphics[width=0.9\columnwidth]{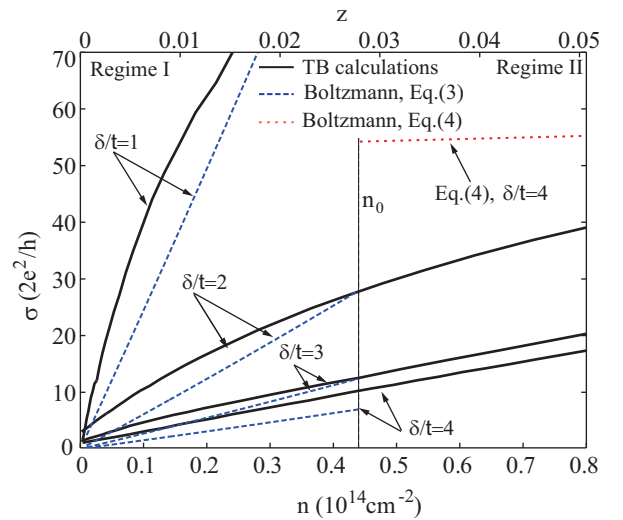}
\end{centering}
\caption{(color online) The density dependence of the TB and Boltzmann
conductivities for Regime I (quantum scattering $z\ll 1$, parabolic band $%
n\ll n_{0}$) and Regime II (quantum scattering $z\ll 1$, linear band $n\gg
n_{0}$); $\protect\xi =a$; $t_{\bot}=0.5t$, corresponding to $%
n_{0}=4.4\times 10^{13}\text{cm}^{-2}$, for selected scattering strengths.
The TB calculations are averaged over 1000 impurity configurations.}
\label{Fig1}
\end{figure}

As shown in the preceding Section, for the case of the Gaussian potential
the Boltzmann approach predicts four different regimes where the density
dependence of the conductivity $\sigma =\sigma (n)$ is qualitatively
different, see Eqns. (\ref{sigma_par}), (\ref{sigma_lin}) and Table I. In
the present Section, we compare these predictions with the exact TB
calculations.

Figure \ref{Fig1} displays the results for Regimes I and II (quantum
scattering, $z\ll 1,$ for the parabolic and linear band, respectively). Let
us start with the most interesting Regime I of the parabolic band
appropriate for most experiments (left part of the figure, $n\ll n_{0})$.
The TB conductivity exhibits a density dependency close to the expected
linear one, $\sigma \propto n$, and the TB and Boltzmann conductivities are
in a reasonable quantitative agreement. Note that for smaller scattering
strengths ($\delta \lesssim 2t)$ the density dependence of $\sigma ^{TB}$ is
rather sublinear, which can be attributed to the quasi-ballistic transport
regime (a transition from the ballistic to diffusive transport regimes in
the SLG is discussed in Refs. \cite{Klos2009,Klos2010}). The calculations
presented in Fig. \ref{Fig1} correspond to the short-range potential with $%
\xi =a$. Similar results are also obtained for a potential with a longer
range, $\xi =3a$ (not shown). Because of computational limitations we are
not able to explore BLG conductivity in the Regime I for larger $\xi .$
However we expect that for a given potential strength the agreement between
the Boltzmann and the TB conductivities improves when the potential becomes
smoother, i.e. when $\xi $ increases. A similar linear dependence of the BLG
conductivity for the Gaussian potential in the parameter range $a\lesssim
\xi \lesssim 5a$ was also obtained by Yuan \textit{et al.} \cite%
{Katsnelson2010}.

Regime II of the quantum scattering case, $z\ll 1$, in the linear band
corresponds to the right part of the figure, $n\gg n_{0}$. In this case the
TB and Boltzmann conductivities disagree both qualitatively and
quantitatively: the TB calculations show a linear dependence $\sigma
^{TB}\propto n,$ whereas Boltzmann approach predicts $\sigma ^{Boltz}=$
const. For the case of the linear band the BLG can be effectively regarded
as two weakly interacting SLG sheets. Therefore the SLG conductivity and the
conductivity of the BLG in the linear band are expected to show similar
features. Indeed, the linear density dependence of the TB conductivity as
well as a qualitative disagreement between the TB and Boltzmann approaches
within the Born approximation for $z\ll 1$ have also been found for Gaussian
scatterers for SLG\cite{Klos2010}.

\begin{figure}[tbp]
%[tbp]
\includegraphics[width=0.9\columnwidth]{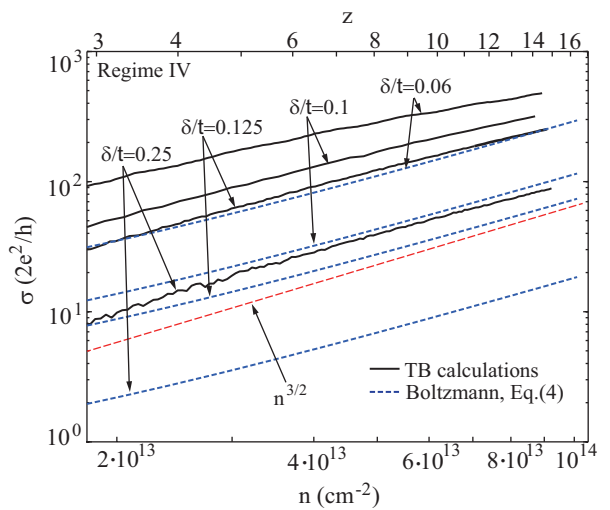}
\caption{(color online) The density dependence of the TB and Boltzmann
conductivities for the Regime IV (classical scattering $z\gg 1$, linear band
$n\gg n_{0}$), $\protect\xi =16a$ ; $t_{\bot}=0.39$ eV, corresponding to $%
n_{0}=2.67\times 10^{12} \text{cm}^{-2}$, for various scattering strengths.
The TB calculations are averaged over 1000 impurity configurations. The
sample size is $L\times W=320\times 20$nm$^{2}$.}
\label{Fig2}
\end{figure}

\begin{figure}[tbp]
%[tbp]
\begin{centering}
\includegraphics[width=0.9\columnwidth]{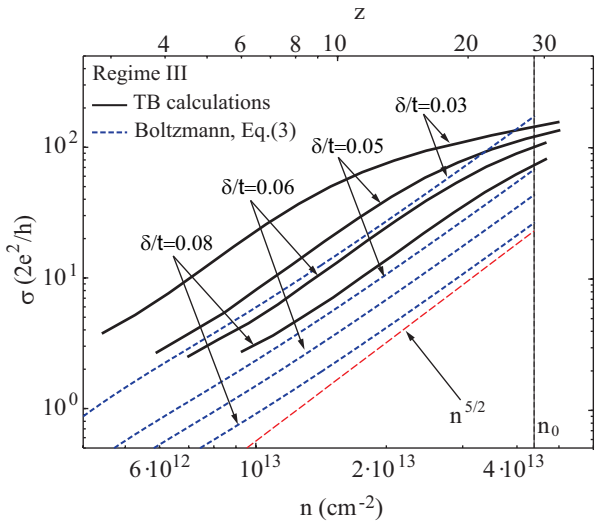}
\end{centering}
\caption{(color online) The density dependence of the TB and Boltzmann
conductivities for the Regime III (parabolic band, $n\ll n_{0}$ and weak
scattering $z\gg 1$). $\protect\xi =32a$; $t_{\bot}=0.5t$ (corresponding to $%
n_{0}=4.4\times 10^{13}\text{cm}^{-2}$), shown for different scattering
strengths. The TB calculations are averaged over 1000 impurity
configurations.}
\label{Fig3}
\end{figure}

Why does the Born approximation for $z\ll 1$ work reasonably well (in the
parameter region explored here) for the BLG in the parabolic band (Regime
I), but fails for BLG in the linear band (Regime II) as well as for the SLG?
The answer to this question can be obtained from an analysis of the validity
of the Born approximation based on non-perturbative approaches\cite%
{Ferreira,priv_com}. Using the $T$-matrix technique for the lattice model or
the partial-wave expansion for the continuum description one can obtain an
exact expression for the scattering rate for the case of the short-range ($%
\delta $-function) impurities. The condition for the validity of the Born
approximation for short-range scattering was discussed by Ferreira\textit{\
et al.}\cite{Ferreira,priv_com} and reads
\begin{subequations}
\label{Born}
\begin{gather}
\frac{2}{\sqrt{\pi }3^{3/4}}\left( V_{eff}/t\right) kR\ln (kR)\ll 1,\text{%
\thinspace SLG;}  \label{Born_SLG} \\
\frac{18t/t_{\perp }}{\left( V_{eff}/t\right) \left( cA_{1}/a^{2}\right) }%
\ll 1,\text{\thinspace BLG (parabolic band)}  \label{Born_BLG}
\end{gather}%
where $k$ is the Fermi wavelength, $R=\frac{3^{3/4}}{2\sqrt{\pi }}a,$ $A_{1}=%
\frac{3\sqrt{3}}{4}a^{2}$ is the area per one C atom, $c$ is a constant of
the order of 1, and $V_{eff}$ is the effective impurity potential obtained
using the $T$-matrix approach and the ab initio calculations\cite%
{Robinson,Wehling2010}. For the SLG (and thus for the BLG in the linear
band), the condition (\ref{Born_SLG}) implies that the Born approximation is
valid only for weak potentials, $V_{eff}/t\ll 5$ . In contrast, condition (%
\ref{Born_BLG}) is more relaxed, such that the Born approximation for the
parabolic band of BLG is expected to work at least qualitatively even for
strong realistic potentials, $V_{eff}/t\sim 60,$ appropriate e.g. for
adsorbed hydrogen\cite{Wehling2010}. In the present study, we present
results for the short range potential with $V_{eff}/t\lesssim 4.$ For larger
$V_{eff}/t$ the structure at hand for a given impurity concentrations and a
system size enters the localization regime, where the Boltzmann approach is
not applicable. Although we expect that the Born approximation would be
qualitatively correct even for such strong potentials, it would be
interesting to explore this case using numerical approaches capable to treat
larger size structures, such as e.g. the Kubo approach\cite{Katsnelson2010}.

Let us now turn to the case of classical scattering, $z\gg 1,$ corresponding
to Regimes III and IV (parabolic and linear bands respectively). We start by
noting that for $z\gg 1$ the conductivity of SLG is well described by the
Boltzmann approach within the Born approximation\cite{Adam2009,Klos2010}. We
therefore expect a similar agreement between the Boltzmann and the TB
calculations for BLG for the case of classical scattering in the linear band
regime (i.e. in Regime IV). The conductivity of BLG in Regime IV
(exemplified by choosing $\xi =16a$) is shown in Fig. \ref{Fig2}. According
to the expectations, the TB calculations are in excellent qualitative
agreement with the Boltzmann predictions showing the expected density
dependence $\sigma \propto n^{3/2}$ and exhibiting the decrease of
conductivity as the impurity strength $\delta $ is increased. However, the
magnitude of the TB conductivity differs from its Boltzmann counterpart by a
factor 3-6 with increasing deviation as $\delta $ increases. A similar
quantitative discrepancy between the TB and the Boltzmann conductivities is
also found for Regimes I and III. One of the reasons for this discrepancy is
that due to computational limitations, the dimension of the structure $L$ is
not sufficiently large to achieve a truly classical diffusive regime with
\end{subequations}
\begin{equation}
L\gg l_{tr}\gg l_{F}  \label{diffusive}
\end{equation}%
where $l_{tr}=v\tau =\frac{h}{4e^{2}}\frac{\sigma }{\sqrt{\pi n}}$ is the
mean free path and $l_{F}$ is the Fermi wavelength. For the scattering
strengths $\delta $ used in calculations typical values of the mean free
path are $l_{tr}\simeq L/2-L/10$ ($L=320$ nm in Fig. \ref{Fig2} and $L=160$
nm for Figs \ref{Fig1},\ref{Fig3}), and $l_{F}\simeq 10$ nm. We expect that
the agreement between the TB and Boltzmann conductivities would improve for
larger structures where the condition Eq. (\ref{diffusive}) is better
satisfied. Another reason for the discrepancy can be related to the
utilization of the Born approximation, the applicability of which improves
as the electron energy increases.

We finally turn to Regime III (classical scattering, $z\gg 1,$ in the
parabolic band). It is exemplified in Fig. \ref{Fig3} by choosing $\xi =32a.$
The TB calculations are in a qualitative agreement with the corresponding
Boltzmann predictions showing the density dependence $\sigma \propto n^{5/2}$
expected for this regime and exhibiting a decrease of conductivity as the
impurity strength $\delta $ is increased. We conclude this discussion by
summarizing that for the classical scattering, $z\gg 1,$ the TB calculations
are consistent with the Boltzmann approach for both parabolic and linear
bands, whereas for the quantum scattering, $z\ll 1,$ the TB and Boltzmann
calculations agree for the parabolic band and disagree for the linear band.

Let us relate our findings to available experimental data. Practically all
the experiments on BLG structures reported so far have been performed for
relatively low electron densities where the dispersion relation is
parabolic. In most of these studies the BLG conductivity exhibits a linear
density dependence. \cite{Morozov2008, Monteverde2010,Xiao2010}. Out of four
regimes considered above, this is consistent only with the Regime I (quantum
scattering in the parabolic band) predicting a linear density dependence for
both short- and long range scatterers. We therefore conclude that Regime I
is the regime corresponding to realistic BLG samples, and the experimentally
observed linear dependence of the BLG conductivity can be caused by both
short- and long-range scattering. Note that exact TB numerical calculations
for SLG for the case of quantum scattering, $z\ll 1,$ also give the same
linear density dependence for both short- and long-range Gaussian scatterers%
\cite{Klos2010}. As the calculated and experimental conductivities of SLG
and BLG are essentially the same, this strongly suggests that the scattering
mechanisms for SLG and BLG can not be different.

\bigskip\ As far as the classical scattering $z\gg 1$ in the parabolic band
is concerned (Regime II), the Boltzmann theory and the TB numerical
calculations predict a superlinear density dependence $\sigma \propto
n^{5/2} $. This dependence has never been observed in BLG structures.\cite%
{Peresreview} This leads us to conclude that the regime of the classical
scattering, $z\gg 1,$ is not accessible experimentally for BLG. (It is
noteworthy that a similar conclusion, namely that the regime of classical
scattering $z\gg 1$ is not achieved in SLG, was reported in Ref.[%
\onlinecite{Klos2010}]). Therefore, the condition $z\equiv \pi n\xi
^{2}\simeq 1$ provides an upper limit for the effective, density dependent
spatial extension of the scatterers present in the experiments. For,
example, for a typical electron density $n=10^{12}$cm$^{-2},$ we obtain $\xi
\simeq 5.6$ nm ($\simeq 22$ lattice constants). This value of the effective
screening length $\xi $ is consistent with the result reported by Ghaznavi
\textit{et al}. \cite{Ghaznavi} for the Poisson-Thomas-Fermi screening of a
charged impurity where a potential drops by a factor of 10 at distances 1-5
nm.

\section{Conclusions}

We have studied the conductivity of BLG utilizing a model of Gaussian
disorder where the effective screening length $\xi $ can be chosen to
describe both short- and long-range scattering. Analytical expressions for
the conductivity of the BLG within the standard Boltzmann approach in the
Born approximation have been derived and compared to exact numerical
tight-binding Landauer-type calculations. Our main findings can be
summarized as follows.

\begin{itemize}
\item The Boltzmann approach for the Gaussian potential predicts four
different regimes where the density dependence of the conductivity $\sigma
=\sigma (n)$ is qualitatively different for parabolic and linear bands ($%
n\ll n_{0}$ respectively $n\gg n${}$_{0}$) and for the quantum and classical
scattering ( $z\ll 1$ respectively $z\gg 1$), see Table I for the summary.
For the classical scattering, $z\gg 1,$ the TB calculations are consistent
with the Boltzmann approach for both parabolic and linear bands, whereas for
the quantum scattering, $z\ll 1,$ the TB and Boltzmann calculations agree
for the parabolic band and disagree for the linear band. We discuss and
analyze the applicability of the Born approximation for BLG and compare it
with the case of SLG.

\item By comparing the obtained density dependencies of the conductivity
with available experimental results we conclude that realistic BLG samples
are in the regime of the parabolic band, $n\ll n_{0},$ and of quantum
scattering, $z\ll 1.$ Here, both short- and long-range scattering lead to
the same linear density dependence of the conductivity. We also compare the
conductivities of SLG and BLG and conclude that they have the same
scattering mechanisms.

\item An upper limit for the effective, density dependent spatial extension
of the scatterers present in the experiments is provided. For, example, for
a typical electron density $n=10^{12}$cm$^{-2},$ we obtain $\xi =5.6$ nm
(corresponding to 22 lattice constants).
\end{itemize}

\begin{acknowledgments}
We are greatly thankful to N. M. R. Peres and A. Ferreira for illuminating
discussions and correspondence and for sending us Ref. \onlinecite{Ferreira} prior
publication. We acknowledge communication with S. Adam at early stages of
this work. I.V.Z. acknowledge support of V.D. and VR.
\end{acknowledgments}

\appendix

\section{Calculation of the Boltzmann conductivity $\protect\sigma $ for
bilayer graphene}

\bigskip In this appendix we present a derivation of the Boltzmann
conductivity for bilayer graphene for the case of the Gaussian potential.
According to Eq. (\ref{sigma}), calculations of the conductivity $\sigma $
requires knowledge of the scattering time $\tau ,$ the DOS at the Fermi
energy $D(E_{F})$, and the Fermi velocity $v_{F}^{BG}$ of the bilayer
graphene.

\subsection{Basics of bilayer graphene}

In this section we briefly summarize basic properties of bilayer graphene
including the dispersion relations, wave functions, DOS and the Fermi
velocity that will be subsequently used below to calculate the conductivity $%
\sigma $ and the scattering time $\tau .$

We write the four component wave function of the bilayer graphene in the
form $\psi =\left( c^{A_{1}},c^{B_{1}},c^{A_{2}},c^{B_{2}}\right) ^{T}e^{i%
\mathbf{Qr}}$ where $A$ and $B$ correspond to the $A$ and $B$ sublattices,
and 1 and 2 to layers, and $\mathbf{Q}$ is the wave vector\cite%
{CastroNetoreview,DasSarmareview}. Substituting this wave function into the
Schr\"{o}dinger equation with the Hamiltonian Eq. (\ref{H}) and expanding $%
\mathbf{Q}$ in the vicinity of the $\mathbf{K}$ point, $\mathbf{Q=K+k}$, we
obtain
\begin{equation}
\begin{pmatrix}
0 & f(\mathbf{q}) & \widetilde{t}_{_{\bot }} & 0 \\
f^{\ast }(\mathbf{q}) & 0 & 0 & 0 \\
\widetilde{t}_{_{\bot }} & 0 & 0 & f^{\ast }(\mathbf{q}) \\
0 & 0 & f(\mathbf{q}) & 0%
\end{pmatrix}%
\begin{pmatrix}
c^{A_{1}} \\
c^{B_{1}} \\
c^{A_{2}} \\
c^{B_{2}}%
\end{pmatrix}%
=-\widetilde{E}%
\begin{pmatrix}
c^{A_{1}} \\
c^{B_{1}} \\
c^{A_{2}} \\
c^{B_{2}}%
\end{pmatrix}%
,  \label{B_eigen_3}
\end{equation}%
where $\widetilde{E}=E/t,$ $\widetilde{t}_{_{\bot }}=t_{\bot }/t,$ and
\begin{eqnarray}
f_{\mathbf{k}} &=&-\frac{3a}{2}\left( k_{x}-ik_{y}\right) =-|f_{\mathbf{k}%
}|e^{-i\theta _{k}};\;  \label{B_f} \\
|f_{\mathbf{k}}| &=&\frac{3a}{2}k=\frac{\hbar v_{F}k}{t};\theta _{k}=\tan
\frac{k_{y}}{k_{x}}.  \notag
\end{eqnarray}%
\newline
The eigenvalues of Eq. (\ref{B_eigen_3}) are easily obtained,%
\begin{equation}
\widetilde{E}_{\mathbf{k}}=s_{1}\left( s_{2}\frac{\widetilde{t}_{_{\bot }}}{2%
}+\sqrt{\frac{\widetilde{t}_{_{\bot }}^{2}}{4}+|f_{\mathbf{k}}|^{2}}\right)
,\;s_{1},s_{2}=\pm 1,  \label{B_eigen_4}
\end{equation}%
with the corresponding eigenfunctions%
\begin{equation}
\psi _{\mathbf{k}}(r)=\frac{1}{\sqrt{C}}%
\begin{pmatrix}
-\widetilde{E}_{\mathbf{k}} \\
f_{\mathbf{k}}^{\ast } \\
s_{1}s_{2}\widetilde{E}_{\mathbf{k}} \\
-s_{1}s_{2}f_{\mathbf{k}}%
\end{pmatrix}%
e^{i\mathbf{kr}},  \label{B_psi_2}
\end{equation}%
where the normalization constant $C=2(\widetilde{E}_{\mathbf{k}}^{2}+|f_{%
\mathbf{k}}|^{2}).$ \newline
%*********************************************************
%                       Fig: bilayer dispersion
%
\begin{figure}[tbp]
\includegraphics[width=0.4\textwidth]{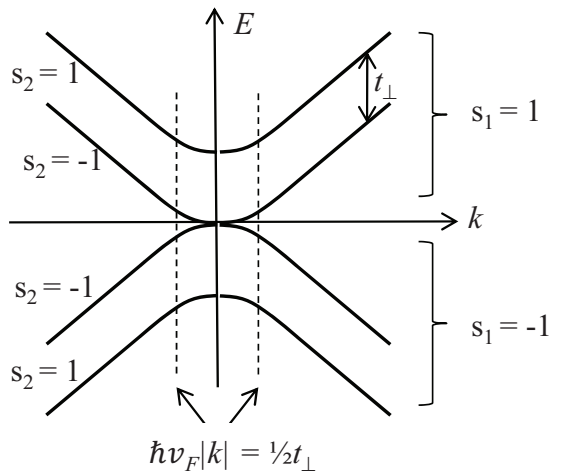} %[tph]
\caption{A schematic illustration of the dispersion relation for a bilayer
graphene, Eq. (\protect\ref{B_eigen_4}).}
\label{band structure}
\end{figure}
%*********************************************************
\newline
The dispersion relation Eq. (\ref{B_eigen_4}) is illustrated in Fig. \ref{band
structure}.

It is instrumental to consider two limiting cases where the dispersion
relation Eq. (\ref{B_eigen_4}) shows qualitatively different behavior, (a) low
energies,  $|f_{\mathbf{k}}|<%
%TCIMACRO{\U{bd}}%
%BeginExpansion
{\frac12}%
%EndExpansion
\widetilde{t}_{\perp }$ ($\hbar v_{F}|k|<%
%TCIMACRO{\U{bd}}%
%BeginExpansion
{\frac12}%
%EndExpansion
t_{\perp })$ and (b) high energies,  $|f_{\mathbf{k}}|<%
%TCIMACRO{\U{bd}}%
%BeginExpansion
{\frac12}%
%EndExpansion
\widetilde{t}_{\perp }$ ($\hbar v_{F}|k|>%
%TCIMACRO{\U{bd}}%
%BeginExpansion
{\frac12}%
%EndExpansion
t_{\perp }).$ Using the relation between the wave number and the electron
density in two dimensional (2D) electron systems, $k=\sqrt{\pi n},$ the
conditions (a) and (b) can be rewritten in terms of the electron density as
follows, (a) low densities, $n\ll n_{0},$ and (b) high densities, $n\gg n${}$%
_{0},$ where $n_{0}=\frac{1}{\pi (3a)^{2}}\left( t_{\bot }/t\right) ^{2}$ is
the critical density separating the parabolic and linear bands. We will
consider only the conduction band ($s_{1}=1$); the same expressions for the
conductivity can be obtained for the valence band ($s_{1}=-1$).

\subsubsection{low densities, $n\ll n_{0}$ (parabolic band)}

Introducing the effective mass $m^{\ast }\equiv \frac{t_{\bot }}{2v_{F}^{2}}$%
, the dispersion relation Eq. (\ref{B_eigen_4}) for the lowest band ($%
s_{2}=-1)$ can be re-written in a form familiar for conventional
semiconductor systems,
\begin{equation}
E=\frac{\hbar ^{2}k^{2}}{2m^{\ast 2}}.  \label{B_low2}
\end{equation}%
Hence, the Fermi velocity for the bilayer graphene $v_{F}^{BG}$ and its DOS
have the same form as the ones for conventional 2D electron systems,
\begin{equation}
v_{F}^{BG}=\hbar k/m^{\ast };\;D^{BG}(E)=\frac{g_{v}g_{s}m^{\ast }}{2\pi
\hbar ^{2}}=\frac{2m^{\ast }}{\pi \hbar ^{2}},  \label{B_DOS_low}
\end{equation}%
where the factors $g_{v}=g_{s}=2$ account for the valley and the spin
degeneracy.

\subsubsection{high densities, $n\gg n_{0}$ (linear bands)}

For the high densities the dispersion relation consists of two linear bands
(as in the case of the the single layer graphene) separated by the energy
interval $t_{\bot }$,

\begin{equation}
E\approx \left\{
\begin{array}{ll}
\hbar v_{F}|k|, & s_{2}=-1; \\
\hbar v_{F}|k|+t_{\bot }, & s_{2}=1.%
\end{array}%
\right.   \label{B_high2}
\end{equation}%
Each of these bands has the same dispersion as the one of single-layer
graphene (SLG); therefore, in this case $v_{F}^{BG}=v_{F},$ and the DOS reads%
\begin{equation}
D^{BG}(E)=2D^{SLG}(E)=\frac{g_{v}g_{s}E}{\pi (\hbar ^{2}v_{F}^{2})}=\frac{4E%
}{\pi (\hbar ^{2}v_{F}^{2})},  \label{B_DOS_high}
\end{equation}%
where a factor of two accounts for two bands.

\subsection{Calculation of the scattering rate}

\label{appendixA} Within the Boltzmann approximation the scattering rate
reads\cite{Ashcroft}
\begin{equation}
\frac{1}{\tau }\mathbf{=}\int \frac{d\mathbf{k}^{\prime }}{(2\pi )^{2}}W_{%
\mathbf{k},\mathbf{k}^{\prime }}(1-\cos \theta ),  \label{A_tau}
\end{equation}%
where in the Born approximation $W_{\mathbf{k},\mathbf{k}^{\prime }}$ is
given by Fermi's Golden Rule,
\begin{equation}
W_{\mathbf{k},\mathbf{k}^{\prime }}=\frac{2\pi }{\hbar }\left\vert V_{%
\mathbf{k},\mathbf{k}^{\prime }}\right\vert ^{2}\delta (E_{k}-E_{k^{\prime
}}),  \label{A_W}
\end{equation}%
with $\theta $ being the angle between the wave vectors $\mathbf{k}$ and $%
\mathbf{k}^{\prime },$ and $V_{\mathbf{k},\mathbf{k}^{\prime }}$ is the
matrix element of the scattering potential $V(\mathbf{r}),$
\begin{equation}
V_{\mathbf{k},\mathbf{k}^{\prime }}=\int \psi _{_{\mathbf{k}}}^{\dagger }V(%
\mathbf{r})\psi _{_{\mathbf{k}^{\prime }}}d\mathbf{r.}  \label{A_V}
\end{equation}%
Using the wave function Eq. (\ref{B_psi_2}) we obtain for the matrix element,%
\begin{equation}
V_{\mathbf{k},\mathbf{k}^{\prime }}=\frac{2\left( \widetilde{E}_{k}%
\widetilde{E}_{k^{\prime }}+f_{k}f_{k^{\prime }}^{\ast }+f_{k}^{\ast
}f_{k^{\prime }}\right) }{2\sqrt{\widetilde{E}_{k}^{2}+|f_{k}|^{2}}\sqrt{%
\widetilde{E}_{k^{\prime }}^{2}+|f_{k^{\prime }}|^{2}}}U_{q},  \label{A_V2}
\end{equation}%
where $U_{q}=\int V(\mathbf{r})e^{i\mathbf{qr}}d\mathbf{r}$ is the Fourier
transform of the scattering potential, and $q=|\mathbf{k-k}^{\prime
}|=2k\sin \frac{\theta }{2}$. We calculate $|U_{q}|^{2}$ making use of the
Wiener-Khinchin theorem\cite{NR} stating that the square modulus of the
Fourier transform of a function is equal to the Fourier transform of its
correlator, $|U_{q}|^{2}=\int e^{i\mathbf{qr}}C(\mathbf{r})d\mathbf{r}$.
Using the expression Eq. (\ref{correlator}) for the correlator of the
Gaussian potential, we get%
\begin{equation}
|U_{q}|^{2}=K(\hbar v_{F})^{2}e^{-\frac{q^{2}\xi ^{2}}{2}}\mathbf{.}
\label{A_WK}
\end{equation}%
One can obtain analytical expressions for the scattering rate Eq. (\ref%
{A_tau}) for two limiting cases of the low densities ($n\ll n_{0}$) and the
high densities ($n\gg n_{0}$) where electrons obey respectively quadratic
and linear dispersion relations.

\subsubsection{low densities, $n\ll n_{0}$ (parabolic band)}

Using Eqns. (\ref{B_f}) and (\ref{B_low2}) in Eq. (\ref{A_V2}) we obtain
\begin{equation}
\left\vert V_{\mathbf{k},\mathbf{k}^{\prime }}\right\vert ^{2}=\frac{1+\cos
2\theta }{2}|U_{q}|^{2},  \label{V_2_BG}
\end{equation}%
The expression for the scattering rate Eq. (\ref{A_tau}) reads,%
\begin{equation}
\frac{1}{\tau }\mathbf{=}\frac{D^{BG}(E)}{4\hbar }\int_{0}^{\pi }d\theta
\left( 1+\cos 2\theta \right) (1-\cos \theta )|U_{q}|^{2},
\label{tau_Gauss_BG}
\end{equation}%
where we used the expression Eq. (\ref{B_DOS_low}) for the DOS. The Fourier
transform of the Gaussian potential is given by the expression Eq. (\ref%
{A_WK}). Substituting it into Eq. (\ref{tau_Gauss_BG}) and performing
integration utilizing that $q=2k\sin \frac{\theta }{2}$, we obtain for the
scattering rate%
\begin{equation}
\frac{1}{\tau }\mathbf{=}\frac{\pi D(E)}{2\hbar }K(\hbar v_{F})^{2}e^{-z}%
\left[ \left( \frac{1}{z}-1\right) I_{1}(z)+\left( \frac{1}{z}+1\right)
I_{2}(z)\right] ,  \label{tau_Gauss_2_BG}
\end{equation}%
where $z=k^{2}\xi ^{2}=\pi n\xi ^{2},$ and $I_{\nu }(z)$ are the modified
Bessel functions of the order $\nu .$ Substituting this expression into Eq. (%
\ref{sigma}) and using the definition of the DOS and $v_{F}^{BG}$ for the
bilayer graphene, Eq. (\ref{B_DOS_low}), we finally obtain Eq. (\ref%
{sigma_par}) for the Boltzmann conductivity.

\subsubsection{high densities, $n\gg n_{0}$ (linear bands)}

Using Eqns. (\ref{B_f}) and (\ref{B_high2}) in Eq. (\ref{A_V2}) we obtain,
\begin{equation}
\left\vert V_{\mathbf{k},\mathbf{k}^{\prime }}\right\vert ^{2}=\frac{\left(
1+\cos \theta \right) ^{2}}{4}|U_{q}|^{2}.  \label{V_2_BG_high}
\end{equation}%
Substituting the expression for into Eq. (\ref{A_tau}) and using Eq. (\ref%
{B_DOS_high}), we obtain for the scattering rate,%
\begin{equation}
\frac{1}{\tau }\mathbf{=}\frac{D^{BG}}{16\hbar }\int_{0}^{\pi }d\theta
\left( 1+\cos \theta \right) ^{2}(1-\cos \theta )|U_{q}|^{2}.
\label{tau_BG_2_high}
\end{equation}%
Substituting Eq. (\ref{A_WK}) into this equation and performing integration
using that $q=2k\sin \frac{\theta }{2}$, we obtain for the scattering rate,%
\begin{equation}
\frac{1}{\tau }\mathbf{=}\frac{\pi D^{BG}}{16\hbar }K(\hbar
v_{F})^{2}e^{-z}\left( \frac{I_{1}(z)+I_{2}(z)}{z}\right) .
\label{tau_Gauss_2_BG_high}
\end{equation}%
Substituting this expression into Eq. (\ref{sigma}) and using the definition
of the DOS and $v_{F}^{BG}$ for the bilayer graphene, Eq. (\ref{B_DOS_high}%
), we finally obtain Eq. (\ref{sigma_lin}) for the Boltzmann conductivity.

%\bibliography{reference}

\begin{thebibliography}{99}
\bibitem{Novoselov2004} K. S. Novoselov, A. K. Geim, S. V. Morozov, D.
Jiang, Y. Zhang, S. V. Dubonos, I. V. Grigorieva, and A. A. Firsov, Science
\textbf{306}, 666 (2004).

\bibitem{CastroNetoreview} A. H. Castro Neto, F. Guinea, N. M. R. Peres, K.
S. Novoselov, and A. K. Geim, Rev. Mod. Phys. \textbf{81, }109 (2009).

\bibitem{DasSarmareview} S. Das Sarma, S. Adam, E. H. Hwang, E. Rossi, Rev.
Mod. Phys., (arXiv:cond-mat/1003.4731).

\bibitem{Peresreview} N. M. R. Peres, Rev. Mod. Phys. \textbf{82}, 2673-2700
(2010)

\bibitem{Novoselov2005} K. S. Novoselov, A. K. Geim, S. V. Morozov, D.
Jiang, M. I. Katsnelson, I. V. Grigorieva, S. V. Dubonos, A. A. Firsov,
Nature \textbf{438}, 197 (2005).

\bibitem{Tan2007} Y.-W. Tan, Y. Zhang, K. Bolotin, Y. Zhao, S. Adam, E. H.
Hwang, S. Das Sarma, H. L. Stormer, and P. Kim, Phys. Rev. Lett. \textbf{99}%
, 246803 (2007).

\bibitem{Morozov2008} S. V. Morozov, K. S. Novoselov, M. I. Katsnelson, F.
Schedin, D. C. Elias, J. A. Jaszczak, and A. K. Geim, Phys. Rev. Lett.
\textbf{100}, 016602 (2008).

\bibitem{Du} X. Du, I. Skachko, A. Barker, and E. Y. Andrei, Nat.
Nanotechology \textbf{3}, 491 (2008).

\bibitem{BolotinPRL} K. I. Bolotin, K. J. Sikes, J. Hone, H. L. Stormer, and
P. Kim, Phys. Rev. Lett. \textbf{101}, 096802 (2008).

\bibitem{Klos2009} J. W. Klos, A. A. Shylau, I. V. Zozoulenko, Hengyi Xu,
and T. Heinzel , Phys. Rev. B \textbf{80}, 245432 (2009).

\bibitem{Ando2006} T. Ando, J. Phys. Soc. Jpn, \text{75}, 074716 (2006).

\bibitem{Nomura2006} K. Nomura and A. H. MacDonlad, Phys. Rev. Lett. \textbf{%
96}, 256602 (2006).

\bibitem{Hwang2007} E. H. Hwang, S. Adam, and S. Das Sarma, Phys. Rev. Lett.
\textbf{98}, 186806 (2007).

\bibitem{Jang2008} C. Jang, S. Adam, J.-H. Chen, E. D. Williams, S. Das
Sarma, and M. S. Fuhrer, Phys. Rev. Lett. \textbf{101}, 146805 (2008).

\bibitem{Zhu2009} W. Zhu, V. Perebeinos, M. Freitag, and Ph. Avouris, Phys.
Rev. B \textbf{80}, 235402 (2009).

\bibitem{Stauber2007} T. Stauber, N. M. R. Peres, and F. Guinea, Phys. Rev.
B \textbf{76}, 205423 (2007).

\bibitem{KatsnelsonNovoselov} M. I. Katsnelson and K. S. Novoselov, Solid
State Commun. \textbf{143}, 3 (2007).

\bibitem{Ostrovsky} P. M. Ostrovsky, I. V. Gornyi, and A. D. Mirlin, Phys.
Rev. B \textbf{74}, 235443 (2006).

\bibitem{Klos2010} J. W. Klos and I. V. Zozoulenko, Phys. Rev. B \textbf{82}%
, 081414 (2010).

\bibitem{Wehling2010} T. O. Wehling, S. Yuan, A. I. Lichtenstein, A. K.
Geim, M. I. Katsnelson, Phys. Rev. Lett. \textbf{105} 056802 (2010).

\bibitem{Ponomarenko2009} L. A. Ponomarenko, R. Yang, T. M. Mohiuddin, M. I.
Katsnelson, K. S. Novoselov, S. V. Morozov, A. A. Zhukov, F. Schedin, E. W.
Hill, and A. K. Geim, Phys. Rev. Lett. \textbf{102}, 206603 (2009).

\bibitem{Monteverde2010} M. Monteverde, C. Ojeda-Aristizabal, R. Weil, K.
Bennaceur, M. Ferrier, S. Gueron, C. Glattli, H. Bouchiat, J.N. Fuchs,
D.Maslov, Phys. Rev. Lett. \textbf{104} 126801 (2010).

\bibitem{Ni2010} Z. H. Ni, L. A. Ponomarenko, R. R. Nair, R. Yang, S.
Anissimova, I. V. Grigorieva, F. Schedin, Z. X. Shen, E. H. Hill, K. S.
Novoselov, A. K. Geim, Nano Lett. \textbf{10}, 3868 (2010).

\bibitem{Xiao2010} S. Xiao, J.-H. Chen, S. Adam, E. D. Williams. and M. S.
Fuhrer, Phys. Rev. B \textbf{82}, 041406 (2010).

\bibitem{Feldman2009} B. Feldman, J. Martin and A. Yacoby, Nat. Phys.
\textbf{5}, 889 (2009).

\bibitem{Adam2008} S. Adam and S. Das Sarma, Phys. Rev. B 77, 115436 (2008).

\bibitem{DasSarma2010} S. Das Sarma, E. H. Hwang, and E. Rossi, Phys. Rev. B
\textbf{81}, 161407(R) (2010).

\bibitem{Katsnelson2007} M. I. Katsnelson, Phys. Rev. B \textbf{76}, 073411
(2007).

\bibitem{Ferreira} A. Ferreira, J. Viana Gomes, J. Nilsson, E. R. Mucciolo,
N. M. R. Peres, A. H. Castro Neto, Phys. Rev. B \textbf{83}, 165402 (2011).

\bibitem{Katsnelson2010} S. Yuan, H. De Raedt, and M. I. Katsnelson, Phys.
Rev. B \textbf{82}, 235409 (2010) % tight-binding

\bibitem{Trushin} M. Trushin, J. Kailasvuori, J. Schliemann, and A. H.
MacDonald, Phys. Rev. B \textbf{82}, 155308 (2010).

\bibitem{Adam2009} S. Adam, P.W. Brouwer, and S. Das Sarma, Phys. Rev. B
\textbf{79}, 201404(R) (2009).

\bibitem{Bardarson} J. H. Bardarson, J. Tworzyd{\l }o, P. W. Brouwer, and C.
W. J. Beenakker, Phys. Rev. Lett. \textbf{99},106801 (2007).

\bibitem{Lewenkopf} C. H. Lewenkopf, E. R. Mucciolo, and A. H. Castro Neto,
Phys. Rev. B \textbf{77}, 081410(R) (2008).

\bibitem{Xu2009} H. Xu, T. Heinzel, and I. V. Zozoulenko, Phys. Rev. B
\textbf{80}, 045308 (2009).

\bibitem{Xu2008} H. Xu, T. Heinzel, M. Evaldsson and I. V. Zozoulenko, Phys.
Rev. B \textbf{77}, 245401 (2008).

\bibitem{Ashcroft} N. W. Ashcroft and N. D. Mermin, \textit{Solid State
Physics}, Holt, Rinehart and Winston, 1976.

\bibitem{priv_com} A. Ferreira and N. M. R. Peres, private communication.

\bibitem{Robinson} J. P. Robinson, H. Schomerus, L. Oroszl\'{a}ny, and V. I.
Fal'ko, Phys. Rev. Lett. 101, 196803 (2008).

\bibitem{Ghaznavi} M. Ghaznavi, Z. L. Mi\v{s}kovi\'{c}, and F. O. Goodman,
Phys. Rev. B \textbf{81}, 085416 (2010).

\bibitem{NR} W. H. Press, S. A. Teukolsky, W. Vetterling, and Brian P.
Flannery, \textit{Numerical Recepies: The Art of scientific Computing},
Cambridge University Press, 2007.
\end{thebibliography}

\end{document}